\documentclass[nofootinbib,showkeys]{revtex4}

\usepackage{latexsym}

\begin{document}

\title{Rainbow metric formalism and Relative Locality}

\author{N. LORET$^*$, L. BARCAROLI$^\dagger$}

\affiliation{Dipartimento di Fisica, Universit\`a di Roma "La Sapienza",\\
and INFN, Sez.~Roma1,\\
P.le A. Moro 2, 00185 Roma, EU.\\
$^\dagger$E-mail: niccolo.loret@uniroma1.infn.it, $^*$E-mail: leonardo.barcaroli@uniroma1.infn.it 
http://www.phys.uniroma1.it}

\begin{abstract}
This proceeding is based on a talk prepared for the XIII Marcell Grossmann meeting. We summarise some results of work in progress in collaboration with Giovanni Amelino-Camelia about momentum dependent (Rainbow) metrics in a Relative Locality framework and we show that this formalism is equivalent to the Hamiltonian formalization of Relative Locality obtained in arXiv:1102.4637. 
\end{abstract}

\keywords{Relative Locality; Rainbow metrics; $\kappa$-Poincar\'{e}; Curved momentum-space.}

\maketitle
\nopagebreak[0]

\section{Introduction}\label{introduzione}
In literature Planck length\footnote{We adopt units such that the speed-of-light scale is $1$ ($c=1$)} $L_P=\sqrt{\hbar G}$ is considered a natural guess as quantum gravity characteristic length scale. However, $L_P$ is non zero only in a regime in which $G$ and $\hbar$ are non zero too, so this hypotesis requires a full fledged quantum gravity theory to elaborate it. We here focus on a "classical non-gravitational" regime \cite{bob, principle, kbob} of quantum gravity. In this regime $\hbar$ and $G$ are both neglected while their ratio is fixed: $M_P = \sqrt{\hbar /G}$. In this paper from now on we will formalize the features of obstruction of measurability and momentum-space deformation with $\kappa$-Poincar\'e Hopf algebra \cite{majruegg,lukruegg} in 1+1D:
\begin{eqnarray}
\{{\cal N},p_0\}=p_i\;,\;\{{\cal N},p_1\}= \left(\frac{1-e^{-2\ell p_0}}{2\ell}-\frac{\ell}{2}p^2\right)\;,\;
\{ p_1,p_0\}=0\;,\label{kAlgebra}
\end{eqnarray} 
where ${\cal N}$ is the boost operator, $P_\alpha$ are the translation operators and where the Casimir operator $\mathcal{C}$ is:
\begin{equation}
{\cal C}=\left(\frac{2}{\ell}\sinh\left(\frac{\ell p_0}{2}\right)\right)^2-p_1^2 e^{\ell p_0}.\label{casimir}
\end{equation}
 The algebric sector of $\kappa$-Poincar\'e can be interpreted as an example of DSR theory\cite{gacIJMPD} with $\ell\sim 1/M_P$ the scale of deformation. It is, therefore, always possible to rely on the existence of a "classical" limit by turning off this deformation ($\ell\rightarrow 0$). In this context moreover it is easier to give a simple interpretation of "Relative Locality" \cite{bob,principle} phenomena on physical observables and thus implement a phenomenology \cite{kbob,transverse}. It is well known in literature (see {\it exempli gratia} \cite{dawn,synchrotron}) that this kind of phenomenology is rather elusive to theoretical investigation. It would be of paramount importance, then, to formalize some kind of {\it rainbow metrics} in relative locality, to try to study Planckian effects in curved spacetimes models. In order to do so we need to define the infinitesimal translation of coordinates between two observers labeled by parameters $\epsilon^\alpha$:
\begin{equation}
\delta x^\beta=\epsilon^\alpha\{p_\alpha,x^\beta\},\label{deftrasl}
\end{equation}
using Poisson brackets to express the action of translation operators on coordinates, as defined in \cite{principle,bob,kbob}. 

\section{Some issues with Rainbow Metrics}\label{RainbowERelLoc}

Smolin and Magueijo \cite{SmolinRainbow} proposed that in a framework where free field theories have plane wave solutions, even though the 4-momentum they carry satisfies deformed dispersion relations of the form
\begin{equation}
p_0^2 f^2\left(p_0/E_P\right)-(p\cdotp p) g^2\left(p_0/E_P\right)=m^2,\label{RainbowSmolin} 
\end{equation} 
spacetime metric should be modified according to the energy of the particle we use to probe it. In fact relation (\ref{RainbowSmolin}) can be realized by the action of a nonlinear map from momentum space to itself, denoted as $U\,:\,\mathcal{P}\rightarrow\mathcal{P}$, given by
$$
U\cdotp(p_0,p_i)=(U_0,U_i)=\left(f\left(p_0/E_P\right)p_0,g\left(p_0/E_P\right) p_i\right),
$$
which implies that momentum space has a nonlinear norm, given by $|p|^2\,=\,\eta^{ab}U_a(p)U_b(p)$. If one still wants to have at his disposal a plane wave solution for free fields, since momentum transforms nonlinearly, the contraction between position and momentum must remain linear. Smolin and Magueijo suggested that in case momentum transforms nonlinearly, this can be obtained imposing that $\zeta^{\alpha\gamma}\tilde{g}_{\gamma\beta}=\delta_\beta^\alpha$, where $\zeta^{\alpha\gamma}$ is the metric of momentum space and $\tilde{g}_{\gamma\beta}$ is the so called "Rainbow metric", such as the spacetime interval
\begin{equation}
ds^2=\tilde{g}_{\gamma\beta}dx^\gamma dx^\beta=(dx^0)^2/f^2-(dx^i)^2/g^2\, ,
\end{equation}
is explicitely energy-dependent. This approach, althought extremely valuable from the phenomenologic point of view, is problematic in case one is interested in avoiding to break the line-element invariance. Let's consider, for example the first order expansion of the dispersion relation we can obtain from casimir (\ref{casimir}):
\begin{equation}
m^2=p_0^2-p_1^2-\ell p_1^2 p_0\,.\label{mdr}
\end{equation}
The mass of (\ref{mdr}) is clearly invariant under a boost generator of the form
\begin{equation}
{\cal N}= x^0 p_1 + x^1\left(p_0-\ell p_0^2 -\frac{\ell}{2}p_1^2\right)\,,\label{boostfirstord}
\end{equation}
obtained imposing $\{{\cal N},{\cal C}\}=0$, assuming $\{p_\alpha,x^\beta\}=\delta_\alpha^\beta$. On the other hand, from equation (\ref{mdr}) we can set
\begin{equation}
f^2(p_0)=1\;,\;\;\; g^2(p_0)=1+\ell p_0\,,
\end{equation}
and then may identify the element line in the flat spacetime case to be
\begin{equation}
ds^2= (dx^0)^2 - (1-\ell p_0)(dx^1)^2\,,
\end{equation}
which is not invariant under the $\ell$-deformed boost (\ref{boostfirstord}).

\section{Metric formalism in Relative Locality}

We will show that is possible to obtain a similar scenario for the $\kappa$-Minkowski spacetime framework in which coordinates satisfy the relation $\{\chi^1,\chi^0\}=\ell \chi^1$, and its deformed symmetry generators algebra $\kappa$-Poincaré time-to-the-right basis defined in Eq. (\ref{kAlgebra}).
Those relations and operators agree with a deformed symplectic sector:
\begin{equation}
\begin{array}{ll}
\{p_0,\chi^0\}=1\, ,&\,\{p_1,\chi^0\}=-\ell p_i\, ,\\
 \{p_0,\chi^1\}=0\, ,&\,\{p_1,\chi^1\}=1\, .
\end{array}
\label{settoresimplettico}
\end{equation}
We can easily obtain the metric formalism as generalization of Pitagora's theorem, expressing the action of translations with Poisson brakets as in Eq.(\ref{deftrasl}):
\begin{equation}
ds^2\equiv  (\epsilon^\alpha \pi_\alpha\rhd\xi^\gamma)(\epsilon^\beta \pi_\beta\rhd\xi_\gamma)=\{p_\kappa,\chi^\mu\}\{p_\lambda,\chi^\nu\}\frac{\partial \pi_\alpha}{\partial p_\kappa}\frac{\partial\xi^\gamma}{\partial \chi^\mu}\frac{\partial \pi_\beta}{\partial p_\lambda}\frac{\partial\xi_\gamma}{\partial \chi^\nu}d\chi^\alpha d\chi^\beta\label{pitagora1}\,, 
\end{equation}
in which $(\xi,\pi)$ is the set of coordinates used by an observer at rest with respect with the center of mass of a certain process, while $(\chi,p)$ is the set used by a generic one. In the $\ell\rightarrow 0$ limit, in which $\{p_\alpha,\chi^\mu\}\equiv\{p_\alpha,x^\mu\}=\delta_\alpha^\mu$, the (\ref{pitagora1}) reduces to
\begin{equation}
ds^2=\eta_{\alpha\beta}e^\gamma_\alpha e^\delta_\beta dx^\alpha dx^\beta
\end{equation} 
where the tetrads are defined as $e^\gamma_\alpha(x)\equiv\partial\xi^\gamma/\partial x^\alpha$. In our case, however we have to take into account the curvature of momentum-space contribution, then $e^\gamma_\alpha(\chi,p)\equiv\frac{\partial\xi^\gamma}{\partial \chi^\mu}\frac{\partial \pi_\alpha}{\partial p_\mu}$. Using (\ref{pitagora1}) it is also possible to find a metric implementing the inferences of momentum-space curvature (also known as Relative Locality effects) on the particle localization process. In fact Eqs. (\ref{settoresimplettico}) and (\ref{pitagora1}) define a momentum dipendent (rainbow) metric, at all orders in $\ell$, of the form: 
\begin{equation}
\tilde{g}_{\alpha\beta}(\chi,p)=\left(\begin{array}{cc}
e^0_0(\chi,p)^2 & -\ell p_1 e^0_0(\chi,p)^2\\
-\ell p_1 e^0_0(\chi,p)^2 & -(e^1_1(\chi,p)^2-\ell^2 p_1^2 e^0_0(\chi,p)^2)
\end{array}\right),\label{mink1+1}\nonumber
\end{equation}
which generates an invariant line-element \cite{exploring}. The minkowskian limit $\tilde{\eta}_{\alpha\beta}(p)$ of (\ref{mink1+1}) is the result of the inference of a deSitter-like curvature of momentum-space on flat spacetime, formalized with the algebra of simmetries we showed in (\ref{kAlgebra}). In this regime, using the geodesic equation for a photon $\tilde{\eta}_{\alpha\beta}\dot{x}^\alpha\dot{x}^\beta=0$, one can finally find the relation  
\begin{equation}
1-2\ell p_1 \frac{d\chi^1}{d\chi^0}-(e^1_1(p)^2-\ell^2 (p_1)^2)\left(\frac{d\chi^1}{d\chi^0}\right)^2=0
\end{equation}
This equation, using the dispersion relation we can obtain from Eq. (\ref{casimir}), gives the expression of a wordline (in terms of commutative coordinates $x^\alpha$) for a massless particle:
\begin{equation}
x^1-\bar{x}^1=-e^{\ell p_0}(x^0-\bar{x}^0),
\end{equation}
which is the same result defined in Ref. \cite{kbob,lateshift} for momentum-dependant massless wordlines. Rainbow metrics formalism is then equivalent to the Hamiltonian one in the spacetime Minkowskian limit of Relative Locality, but it can be even more useful in phenomenology since it naturally implements spacetime curvature. Further analyses should be dedicated to the comparison between this approach and the promising investigation of Relative Locality in curved spacetimes described in \cite{lateshift,jackcurvo, finsler}.

\end{document}